\documentclass[11pt]{llncs}
\usepackage{graphicx}
\usepackage{cite}
\usepackage{url}
\long\def\omit#1{\relax}    % MARK TEXT TO BE INCLUDED ONLY IN LONG VERSION

\usepackage{times}
\usepackage{mathptm}

\usepackage{amssymb}
\usepackage{color}
\usepackage{algorithmic}
\usepackage[ruled]{algorithm}

\newtheorem{thm}{Theorem}
\newtheorem{cor}[thm]{Corollary}
\newtheorem{lem}[thm]{Lemma}

\newtheorem{defn}[thm]{Definition}

\newtheorem{clm}[thm]{Claim}

\let\oldendproof\endproof
\def\endproof{\qed\oldendproof}

\newcommand{\Real}{\mathbb R}
\newcommand{\e}{\epsilon}

\newcommand{\A}{\mathcal{A}}
\newcommand{\R}{\mathcal{R}}
\newcommand{\M}{\mathcal{M}}

\begin{document}

\title{Approximate Weighted Farthest Neighbors\\ and Minimum Dilation Stars}

\author{{\large John Augustine, David Eppstein, and Kevin A. Wortman}}

\institute{Computer Science Department\\
University of California, Irvine\\
Irvine, CA 92697, USA\\
{\tt \{jea,eppstein,kwortman\}@ics.uci.edu}}

\maketitle

\pagestyle{plain}
\thispagestyle{empty}

\begin{abstract}
We provide an efficient reduction from the problem of querying approximate multiplicatively weighted farthest neighbors in a metric space to the unweighted problem.  Combining our techniques with core-sets for approximate unweighted farthest neighbors, we show how to find $(1+\epsilon)$-approximate farthest neighbors in time $O(\log n)$ per query in $D$-dimensional Euclidean space for any constants $D$ and $\epsilon$.  As an application, we find an $O(n \log n)$ expected time algorithm for choosing the center of a star topology network connecting a given set of points, so as to approximately minimize the maximum dilation between any pair of points.
\end{abstract}

\section{Introduction}

Data structures for proximity problems such as finding closest or farthest neighbors or maintaining closest or farthest pairs in sets of points have been a central topic in computational geometry for a long time~\cite{am05}.  Due to the difficulty of solving these problems exactly in high dimensions, there has been much work on approximate versions of these problems, in which we seek neighbors whose distance is within a $(1+\epsilon)$ factor of optimal~\cite{dg05}.  In this paper, we consider the version of this problem in which we seek to approximately answer farthest neighbor queries for point sets with multiplicative weights.  That is, we have a set of points $p_i$, each with a weight $w(p_i)$, and for any query point $q$ we seek to approximate $\max_i w(p_i)\,d(p_i,q)$.

We provide the following new results:

\begin{itemize}
\item We describe in Theorem~\ref{thm:general} a general reduction from the approximate weighted farthest neighbor query problem to the approximate unweighted farthest neighbor query problem.  In any metric space, suppose that there exists a data structure that can answer unweighted $(1+\epsilon)$-approximate farthest neighbor queries for a given $n$-item point set in
query time $Q(n,\epsilon)$, space $S(n,\epsilon)$, and preprocessing time $P(n,\epsilon)$.  Then our reduction provides a data structure for answering $(1+\epsilon)$-approximate weighted farthest neighbor queries in time $O(\log n + ({1\over\epsilon}\log{1\over\epsilon})Q(n,\epsilon/2))$ per query,
that uses space $O(n+ ({1\over\epsilon}\log{1\over\epsilon})S(n,\epsilon/2))$, and can be constructed in preprocessing time
$O(n\log n+({1\over\epsilon}\log{1\over\epsilon})P(n,\epsilon/2))$.

\item We apply core-sets~\cite{AgaHarVar04} to find a data structure for the approximate unweighted farthest neighbor query problem in $\Real^D$,  for any constant $D$, with query time $O(\e^{\frac{1-D}{2}})$, space $O(\e^{\frac{1-D}{2}})$, and preprocessing time  $O(n + \e^{\frac32-D})$ (Theorem~\ref{thm:euclid-u}).  Applying our reduction results in a data structure for the approximate weighted farthest neighbor query problem in $\Real^D$, with preprocessing time  $O(n \log n + \e^{\frac12-D}\log\frac{1}{\e})$, query time $O(\log n + \e^{\frac{-D-1}{2}}\log\frac{1}{\e})$, and space $O(n)$ (Corollary~\ref{cor:euclid-w}).

\item As a motivating example for our data structures, we consider the problem of finding
a star-topology network for a set of points in $\Real^D$, having one of the input points as the hub of the network, and minimizing the maximum dilation of any network path between any pair of points.
By results of Eppstein and Wortman~\cite{ew05}, this problem can be solved exactly in time
$O(n\,2^{\alpha(n)}\log^2 n)$ in the plane, and $O(n^2)$ in any higher dimension.  By using our data structure for approximate weighted farthest neighbor queries, we find in Corollary~\ref{cor:mds-hub} a solution to this problem having dilation within a $(1+\epsilon)$ factor of optimal, for any constant $\epsilon$ and constant dimension, in expected time $O(n\log n)$.  More generally, as shown in Theorem~\ref{thm:mds-punchline}, we can approximately evaluate the dilations that would be achieved by using every input point as the hub of a star topology, in expected $O(n\log n)$ total time.
\end{itemize}

\section{Problem Definition}

We first define the \emph{weighted farthest neighbor query problem} in a metric space, and then extend its definition to relevant variants and special cases. Let $\M = (S, d)$ be a metric space.   For any two points $p$ and $q$ in $S$, $d(p,q)$ denotes the distance between them.  We assume we are given as input a finite set $P\subset S$; we denote by $p_i$ the points in $P$.  In addition, we are given positive weights on each point of $P$, which we denote by a function $w:P\mapsto \Real^+$.
%In addition, we are given a positive weight $w_i$ associated with each point $p_i$.  
We wish to preprocess the point set such that for any query point $q \in S$, we can quickly produce a point $f \in P$ that maximizes the weighted distance from $q$ to $r$.  More precisely, $f$ should be the point in $P$ that maximizes (or approximately maximizes) the weighted distance $d_w(q,p_i) = w(p_i) \cdot d(q,p_i)$ from the query point~$q$. If we restrict the weights of all points to be equal, then this problem simplifies to the \emph{unweighted farthest neighbor query problem}.

Finding the exact farthest neighbor to a query point can be computationally expensive. Hence, we are interested in approximate solutions to this problem, with sufficient guarantees that our algorithm produces a reasonably distant point in the weighted sense. To achieve this, we define the \emph{approximate weighted farthest neighbor query problem}, wherein we are given an input of the same type as the weighted farthest neighbor query problem, and wish to preprocess the point set so that we can efficiently produce an $\e$-approximate weighted  farthest point.  That is, we wish to find a point $r \in P$ such that $d_w(q,r) \geq (1-\e) d_w(q,p_i)$ for some predefined $\e$ and all $p_i\in P$.  A similar definition holds for the \emph{approximate unweighted farthest neighbor query problem} as well. When the metric space under consideration is Euclidean, we call the problem the \emph{Euclidean approximate weighted farthest neighbor query problem}. 

Without loss of generality, we can assume that $0 < w(p) \leq 1$ for all $p \in P$, and that exactly one point $p_1$ has weight $1$.  The first assumption can be made to hold by dividing all weights by the largest weight; this leaves approximations to the weighted farthest neighbor unchanged.  The second assumption can be made to hold by arbitrarily breaking ties among contenders for the maximum-weight point.

\section{Reduction from Weighted to Unweighted}

First, let us suppose that we have a family of  algorithms $\A_\e$ for the  approximate unweighted farthest neighbor query problem defined over any metric space $\M = (S, d)$. $\A_{\e}$ consists of two components, one for preprocessing and the other for querying. For an input instance of the approximate unweighted farthest neighbor query problem,  let $f \in P$ maximize the distance between our query point $q$ and $f$. Using $\A_{\e}$, we can  find an $\e$-approximate farthest point $r$ such that $(1-\e) d(q,f) \leq d(q,r)$. The running time of $\A_{\e}$ has two components, namely preprocessing time and query time denoted $P(n, \e)$, and $Q(n, \e)$ respectively. It has a space requirement of $S(n, \e)$. Our real concern is the weighted version of the problem and in this section, we provide a data structure to reduce an instance of the approximate weighted farthest neighbor query problem such that it can be solved by invoking $A_\e$. We call this family of reduction algorithms $\R_\e$.

The preprocessing and querying components of $\R_\e$ are provided in Procedures \ref{alg:reduction} and \ref{alg:query} respectively. 
%Let $\e = \frac{2 \ep}{(1+\ep)}$. 
For convenience, we separate out the points in $S$ into two subsets $S_1$ and $S_2$. The first subset $S_1$ contains  all points in $P$ with weights in $[\e/2, 1]$, and $S_2$ contains $P \setminus S_1$.  We use different approaches for the two subsets producing three candidate approximate farthest neighbors. Our final result is the farther of the three candidates in the weighted sense. We now show that the algorithm is correct both when $f \in S_1$ and when $f \in S_2$.

\subsection*{Case $f \in S_1$: } In this case, we only consider points in $S_1$. Therefore, we can assume that the weights are within a factor of $2/\e$ of each other. We partition the points set $S_1$ into buckets such that each bucket $B_i$ contains all points in $S_1$ with weight in $[(\e/2) (1+\e/2)^{(i-1)}, (\e/2) (1+\e/2)^i)$.
The number of buckets needed to cover $S_1$ is $\Theta((1/\e) \ln(1/\e))$.

% ADD THIS BACK IN THE JOURNAL VERSION
\omit{
The following claim shows that  the number of buckets needed to cover $S_1$ is at most $(0.5 + 2/\e) \ln (2/\e) \in \Theta((1/\e) \ln(1/\e))$.

\begin{clm}
\label{clm:numbuckets}
For all $x > 0$, $x (1+x)^{((0.5 + 1/x) \ln (1/x))} \geq 1$ 
\end{clm}
\begin{proof}
% I don't know the best way to introduce the following inequality taken from Motwani et al. Ask David about it.
\begin{eqnarray*}
(1+x)^{(0.5 + 1/x)} &\geq& e \\ % taken from prop 3 on pg 435 of Randomized Algorithms by Motwani and Raghavan
\Rightarrow \quad ((1+x)^{(0.5 + 1/x)})^{\ln (1/x)} &\geq& e^{\ln (1/x)} = 1/x\\
%\Rightarrow \quad  (1+x)^{((0.5 + 1/x)\ln 1/x)} &\geq& 1/x \\
\Rightarrow \quad  x (1+x)^{((0.5 +1/x) \ln (1/x))} &\geq& 1
\end{eqnarray*}
\end{proof}
}% END OMITTED PROOF

Our data structure consists of separate instances (consisting of both the preprocessing and querying components) of $A_\e$ with the error parameter $\e/2$ for each bucket $B_i$. In addition, each $B_i$ is preprocessed by the preprocessing instance of $A_\e$. To query our data structure for the $(\e/2)$-approximate farthest neighbor $k_i \in B_i$ from $q$, we run the querying instance of $\A_\e$ on each preprocessed bucket $B_i$. Our candidate approximate weighted farthest neighbor for this case will be the $r_1 = \arg\max_{k_i} d(q, k_i)$. 
\begin{lem}
\label{lem:inside}
If $f \in S_1$, then
\begin{equation}\label{eqn:R1}
d_w(q,r_1) \geq (1 - \e) d_w(q, f),
\end{equation}
where $f$ is the exact weighted farthest point and $r_1$ is as defined above.
\end{lem}
\begin{proof}
%Note that $\ep = \frac{\e}{2-\e}$. 
Let us just consider bucket $B_i$ whose weights are between some  $(\e/2) (1+\e/2)^{(i-1)}$ and $(\e/2) (1+\e/2)^i$. 
%$C_i$ is the $\ep$-kernel for this bucket. 
Let $f_i$  be the exact weighted farthest point from $q$ in $B_i$. Within the bucket $B_i$, $k_i$ is an $(\e/2)$-approximate farthest point if we don't consider the weights. I.e., 
\[
(1-\e/2) d(q, f_i) \leq d(q, k_i).
\]
However, when we consider the weighted distance, $w(k_i)$ and $w(f_i)$ differ by a factor of at most $(1+\e/2)$
%we can be off by a factor of at most $1/(1+\ep)$ 
and hence our inequality for the weighted case becomes
\[
\frac{1-\e/2}{1+\e/2} d_w(q, f_i) \leq d_w(q, k_i).
\]
Since $\frac{1-\e/2}{1+\e/2} > (1-\e)$ for $\e > 0$, we get 
\begin{equation}
\label{eqn:approx}
(1-\e) d_w(q, f_i) \leq d_w(q, k_i).
\end{equation}

Let the bucket $B_{i^\ast}$ contain the weighted farthest point $f$. By definition, $d_w(q, r_1) \geq d_w(q, k_{i^\ast})$, which proves the lemma when combined with Equation \ref{eqn:approx}.
\end{proof}

\floatname{algorithm}{Procedure}
\begin{algorithm}[p]
\caption{Preprocessing for the approximate unweighted farthest neighbor problem}
\label{alg:reduction}
\begin{algorithmic}[1]

\STATE{\bf INPUT:} Set of points $P$ with $n= |P|$ and a distance metric $d$. A fixed error parameter $\e$.

\COMMENT{Farthest point is in $S_1$.}
\STATE {\bf Let} $S_1 \subseteq P$ be points  with weights in $[2/\e, 1]$. 

\STATE {\bf Partition} $S_1$ into buckets $B_0, B_1, \ldots$ such that each $B_i$ contains all points in $S_1$ whose weights are  in $((\e/2) (1+\e/2)^{i-1}, (\e/2) (1+(\e/2))^{i}]$.

\STATE {\bf Preprocess} each bucket by calling the preprocessing component of $A_\e$ instance with error parameter $\e/2$.

\COMMENT{Farthest point is in $S_2$.}
\STATE {\bf Assign} $p_1$ to be point in $P$ with highest weight (assumed to be 1).

\STATE {\bf Sort} $S_2 = P \setminus S_1$ in non-decreasing order of $d(p_1,s)$ for all $s \in S_2$.

\FOR{ each $s$ taken in reverse sorted order from $S_2$}
\IF{$\exists x \succ s$ in $S_2$ such that $d_w(p_1,s) \leq d_w(p_1, x)$}
\STATE $S_2 \leftarrow S_2 \setminus \{s\}$. \\
\COMMENT{Note that this can be done in $O(n)$ time by updating the $x$ that maximizes $d_w(p_1, x)$ at each iteration.}
\ENDIF
\ENDFOR
\STATE {\bf Ensure} that $S_2$ is stored in a manner suitable for binary searching.
\end{algorithmic}
\end{algorithm}

\begin{algorithm}[p]
\caption{Query step of the approximate unweighted farthest neighbor problem}
\label{alg:query}
\begin{algorithmic}[1]

\STATE{\bf INPUT:} Query point $q$, $B_i$ for all $1 \leq i \leq 1/\e \ln(1/\e)$, and $S_2$.

\COMMENT{Farthest point is in $S_1$.}
\STATE {\bf Assign} $r_1 \leftarrow q$

\FOR{ each $B_i$}

\STATE $k_i \leftarrow \A_{(\e/2)}(B_i)$

\IF{$d_w(q, r_1) \leq d_w(q, k_i)$}
     \STATE $r_1 \leftarrow k_i$
\ENDIF
\ENDFOR

\COMMENT{Farthest point is in $S_2$.}
\STATE {\bf Binary Search} through $S_2$ for point $r_2$ such that $d(p_1, r_2) \geq d(p_1, q)/\e$ and $d_w(p_1, r_2)$ is maximized.

\STATE {\bf Return} $r \leftarrow \arg\max_{\{r_1, r_2\, p_1\}} (d_w(q, r_1), d_w(q, r_2), d_w(q, p_1)) $ 

\end{algorithmic}
\end{algorithm}

\subsection*{Case $f \in S_2$: }

We sort the points in $S_2$ in non-decreasing order of their distance from $p_1$; recall that $p_1$ is the only point in $P$ with $w(p_1) = 1$. We then pare down $S_2$ by preserving points $x$ such that $d_w(p_1,x) > d_w(p_1,y)$ for all $y \succ x$ in the sorted sequence of points, and discarding the rest. Our data structure for this case consists of the remaining reduced sorted list, stored in a manner suitable for performing binary searches (e.g. an array or binary search tree). To query the data structure, we binary search for the point $r_2$ that is the first point in the reduced sorted $S_2$ farther than $d(p_1, q)/\e$ from $p_1$. Our two candidates for this case are $r_2$ and $p_1$.

\begin{lem}\label{lem:outside}
If $f \in S_2$, and $p_1$ is not an $\e$-approximate weighted farthest point from $q$, 
%and $r_2$ is as defined above,
then $d_w(q,r_2) \geq (1-\e) d_w(q,f)$.
\end{lem}
\begin{proof}
Consider the ball whose center is $p_1$ and radius is $2 d(q,p_1)/ \e$. Since $f \in S_2$, we can say that $d(q,f) > 2 d(q,p_1) / \e$. Otherwise, $d_w(q,f) \leq (\e/2) d(q, f) \leq d(q,p_1) = d_w(q, p_1)$ (since $w(p_1) = 1$) and this contradicts our assumption that $f \in S_2$. In other words, both $f$ and $r_2$ fall outside this ball, the former by argument and the latter by construction. Hence, 
\begin{equation}
d(p_1, r_2) \geq 2 d(p_1,q)/ \e. \label{eqn:radR}  
\end{equation}
In addition, $r_2$ is chosen over $f$ as being the farthest weighted neighbor of $p_1$ because  (by construction) it is the weighted farthest point from $p_1$ outside the ball in consideration. Hence,
\begin{equation}\label{eqn:radF}
    d_w(p_1,r_2) \geq d_w(p_1,f).
\end{equation}
Consider $\triangle p_1qr_2$.
\begin{eqnarray*}
d(p_1, r_2) &\leq& d(p_1, q) + d(q, r_2) \\
&\leq& (\e/2) d(p_1, r_2) + d(q, r_2) \quad \quad \mbox{(due to \ref{eqn:radR})}\\
\Rightarrow \quad d(q, r_2) &\geq& (1-\e/2) d(p_1, r_2).
\end{eqnarray*}
Again,
\begin{eqnarray*}
d(p_1, r_2) + d(p_1, q) &\geq& d(q, r_2) \\
d(p_1, r_2) + (\e/2) d(p_1, r_2) &\geq& d(q, r_2)  \quad \quad \mbox{(due to \ref{eqn:radR})}\\
(1+\e/2)d(p_1, r_2) &\geq& d(q, r_2).
\end{eqnarray*}
Therefore,
\begin{equation}\label{eqn:one}
    (1-\e/2) d(p_1, r_2) \leq d(q, r_2) \leq (1+\e/2) d(p_1, r_2).
\end{equation}
With similar treatment of $\triangle p_1qf$, we get 
\begin{equation}\label{eqn:two}
    (1-\e/2) d(p_1, f) \leq d(q, f) \leq (1+\e/2) d(p_1, f).
\end{equation}
From Equation \ref{eqn:two}, we have 
\begin{eqnarray*}
d_w(q,f) &\leq& (1 + \e/2) d_w(p_1,f) \\
&\leq&(1+\e/2) d_w(p_1,r_2) \quad \quad \mbox{(From \ref{eqn:radF})}\\
&\leq&\left(\frac{1+\e/2}{1-\e/2}\right) d_w(q, r_2) \quad \quad \mbox{(From \ref{eqn:one})}\\
\Rightarrow \quad d_w(q, r_2) &\geq& (1 - \e) d_w(q,f).
\end{eqnarray*}
\end{proof}

\begin{thm}
\label{thm:general}
The family of reduction algorithms $\R_{\e}$ answer $(1+\e)$-approximate weighted farthest neighbor queries in time $O(\log n + (\frac{1}{\epsilon}\log{1\over\epsilon})Q(n,\e/2))$ per query, use space $O(n+ ({1\over\e}\log{1\over\e})S(n,\e/2))$, and the data structure can be constructed in preprocessing time $O(n\log n+({1\over\e} \log{1\over\e})P(n,\e/2))$. 
\end{thm}
\begin{proof}
The proof of approximation guarantee follows from Lemmas \ref{lem:inside} and \ref{lem:outside} in a straightforward manner. Preprocessing takes time $O(({1\over\e} \log{1\over\e})P(n,\e/2))$ for partitioning into buckets and calling the preprocessing component of $\A_\e$, and $O(n \log n)$ for sorting the points, hence accounting for the total preprocessing  time. While querying, we call the querying component of $A_\e$ for each bucket costing us $O((\frac{1}{\epsilon}\log{1\over\epsilon})Q(n,\e/2))$, and the binary search takes time $O(\log n)$ adding to the total querying time. Similarly, the space taken to store the preprocessed buckets and the sorted sequence of points is $O(({1\over\e}\log{1\over\e})S(n,\e/2))$ and $O(n)$ respectively, adding to the total space required.

\end{proof}

The preprocessing and querying components (Procedures \ref{alg:reduction} and \ref{alg:query} respectively) are presented in pseudocode format. Each procedure has two parts to it corresponding to the two cases when $f \in S_1$ and $f \in S_2$. In Procedure \ref{alg:reduction}, we first partition $S_1$ into buckets and preprocess them. Secondly, we sort $S_2$ and pare it down according to the requirements of Lemma \ref{lem:outside}. In Procedure \ref{alg:query}, we choose the farthest neighbor $r_1$ of $q$ in $S_1$ by querying each bucket and choosing the farthest from the pool of results. Secondly, we binary search in the reduced $S_2$ to obtain the point $r_2$ that is at or just farther than a distance of  $2 d(q, p_1)/\e$ from $q$. Our final $\e$-approximate farthest neighbor $r$ of $q$ is the weighted farthest point in $\{r_1, r_2, p_1\}$ from $q$.

\section{Euclidean Approximate Unweighted Farthest Neighbor Queries}

\label{data-structure}

In this section, our problem definition remains intact except for the restriction that $S$ is  $\Real^D$, where $D$ is a constant, with the Euclidean distance metric. Notice that our points are unweighted and we are seeking a family of approximation algorithms $\A_\e$ that we assumed to exist as a black box in the previous section. Let $f \in P$ maximize the distance between our query point $q$ and points in $S$. For a given $\e$, we are asked to find an $\e$-approximate farthest point $r$ such that $d(q,r) \geq (1-\e) d(q,f)$. We can solve this using the $\e$-kernel technique surveyed in \cite{AgaHarVar04}. Let $u$ be the unit vector in some direction. We denote the directional width of $P$ in direction $u$ by $w(u, P)$ and is given by
\[
w(u,P) = \max_{s \in P} \langle u, s \rangle - \min_{s \in P} \langle  u, s \rangle,
\]
where $\langle u, s \rangle$ is the dot product of $u$ and $s$. An $\e$-kernel is a subset $K$ of $P$ such that for all unit directions $u$,
\[
(1 - \e) w(u, P) \leq w(u,K).
\]
It is now useful to state a theorem from \cite{Cha04, AgaHarVar04} that provides us an $\e$-kernel in time linear in $n$ and polynomial in $1/\e$ with the dimension $D$ appearing in the exponent.

\begin{thm}
Given a set $P$ of $n$ points in $\Real^D$ and a parameter $\e > 0$, one can compute an $\e$-kernel of $P$ of size $O(1/\e^{(D-1)/2})$ in time $O(n + 1/\e^{D - (3/2)})$.
\end{thm}
Consider points $s_1$ and $s_2$ in $P$ and $k_1$ and $k_2$ in $K$ that maximize the directional widths in the direction  $u$, which is the unit vector in the direction of $\overrightarrow{qf}$. 
\[
(1-\e) w(u, P) = (1-\e) \langle u, s_1 \rangle - (1-\e) \langle u, s_2 \rangle \leq \langle u, k_1 \rangle - \langle u, k_2 \rangle.
\]
Now, $\langle u, k_2 \rangle \geq \langle u, s_2 \rangle$, because otherwise, $w(u, P)$ can be maximized further. Hence with some substitution, we get
\[
(1-\e) \langle u, s_1 \rangle \leq \langle u, k_1 \rangle.
\]

The point $f$  maximizes the left hand side in the direction of $\overrightarrow{qf}$ because it is the farthest point from $q$. Therefore, the above inequality suggests that $k_1 \in K$ is an $\e$-approximate farthest neighbor of $q$. This implies that the point farthest from $q$ obtained by sequentially searching the $\e$-kernel $K$ will be an $\e$-approximate farthest neighbor of $q$. Hence, we can state the following theorem.

\begin{thm}
\label{thm:euclid-u}
Given a set $P$ of $n$ points in $\Real^D$ and a parameter $\e > 0$, there exists a family of algorithms $\A_\e$ that answers  the $\e$-approximate unweighted farthest neighbor query problem in preprocessing time $O(n + e^{\frac32-D})$, and the query time and space requirements are both $O(\e^{\frac{1-D}{2}})$.
\end{thm}

Combining this theorem with our general reduction from the weighted to the unweighted problem gives us the following corollary.

\begin{cor}
\label{cor:euclid-w}
Given a set $P$ of $n$ points in $\Real^D$, for any constant $D$, and given a weight function $w:P\mapsto \Real^+$ and a parameter $\e>0$, there exists a family of algorithms that answers the $\e$-approximate weighted farthest neighbor query problem in preprocessing time $O(n \log n + \e^{\frac12-D}\log\frac{1}{\e})$ and query time $O(\log n + \e^{\frac{-D-1}{2}}\log\frac{1}{\e})$ with a space requirement of $O(n)$.
\end{cor}

\begin{proof}
We apply Theorem~\ref{thm:general} to reduce the weighted problem to a collection of instances of the unweighted problem, which we solve using Theorem~\ref{thm:euclid-u}.  The query time comes from combining the results of these two theorems, and the space bound comes from the fact that our overall data structure consists only of a sequence of points together with a family of disjoint subsets of points forming one core-set for each of the instances of the unweighted problem used by Theorem~\ref{thm:general}.  To calculate the time bound, note that Theorem~\ref{thm:general} takes time $O(n\log n)$ to sort the low-weight points by distance from $p_1$, and in the same time we may also perform the partition of high-weight points into buckets used by that theorem.  Once we have partitioned the points into buckets, we construct for each bucket an instance of the data structure of  Theorem~\ref{thm:euclid-u};
this takes time $O(n_i + e^{\frac32-D})$ per bucket and adding these bounds over all buckets gives the stated preprocessing time bound.
\end{proof}

\section{Constrained Minimum Dilation Stars}

The \emph{dilation} between two vertices $v$ and $w$ of a weighted
graph is defined as the ratio of the weight of the shortest path from
$v$ to $w$, divided by the direct distance between $v$ and $w$.  The
dilation of the entire graph is defined as the greatest dilation
between any pair of vertices in the graph.  A \emph{star} is a
connected graph with exactly one internal vertex, called its
\emph{center}.  Any collection of $n$ points admits $n$ possible
stars, since any individual point may serve as a star center.  In this
section we consider the problem of computing the dilation of all of
these $n$ stars.  A solution to this problem provides the foundation
for a solution to the problem of choosing an optimal center: simply
search for the point whose corresponding dilation value is smallest.

Eppstein and Wortman considered \cite{ew05} the problem of selecting
star centers that are optimal with respect to dilation.  They showed
that for any set of $n$ points $P \subset \mathbb{R}^D$, there exists
a set $C$ of $O(n)$ pairs of points such that the worst pair for any
center $c \in \mathbb{R}^D$ is contained in $C$.  They give an $O(n
\log n)$-time algorithm for constructing $C$, and go on to consider
the problem of computing the center that admits a star with minimal
dilation.  They present an $O(n \log n)$ expected-time algorithm for
the case when $c$ may be any point in $\mathbb{R}^D$ and $D$ is an
arbitrary constant, and an $O(n \, 2^{\alpha(n)} \log^2 n)$
expected-time algorithm for the case when $c$ is constrained to be one
of the input points and $D=2$.

These results imply that the dilation of all $n$ stars with centers
from $P$ may be computed in $O(n^2)$ time: construct $C$, then for
each $c \in P$ and $(v,w) \in C$, evaluate the dilation of the path
$\langle v, c, w \rangle$.  In this section we improve on this time
bound through approximation.  We will show that a
$(1-\epsilon)$-approximation of the dilation of all $n$ stars may be
computed in $O(n \log n)$ expected time, and hence an approximately
optimal center $c \in P$ may be identified in $O(n \log n)$ expected
time.

Our approximation algorithm first uses the results of \cite{ew05} to
compute the optimal $c_{OPT} \in \mathbb{R}^D$.  Note that $c_{OPT}$
may or may not be in $P$, and in general will not be.  Our algorithm
then partitions $P$ into two subsets: the $k \in O(1)$ points nearest
to $c_{OPT}$, and the other $n-k$ points.  We say that a star center
$c \in P$ is \emph{$k$-low} if it is one of the $k$ points nearest to
$c_{OPT}$, or \emph{$k$-high} otherwise.  The dilation values for all
the $k$-low centers are computed exactly in $O(n \log n)$ time using a
combination of known techniques.

The dilation between any two points $p_i, p_j \in P$ through any
$k$-high center $c$ may be approximated by a weighted distance from
$c$ to the centroid of $p_i$ and $p_j$.  Hence the the dilation of the
star centered on any $k$-high $c$ may be approximated by the distance
from $c$ to the centroid farthest from $c$.  We use the data structure
described in Section \ref{data-structure} to answer each of these
$n-k$ weighted farthest-neighbor queries in $O(\log n)$ time, which
makes the overall running time of the approximation algorithm $O(n
\log n)$.

\begin{defn}
\label{defn:dilation}
Let $G$ be a Euclidean star with vertices $V$ and center $c$.  The
dilation between any $v, w \in V \setminus \{c\}$ is

\[ \delta(c,v,w) = \frac{|v c| + |w c|}{|v w|} \]

\noindent and the dilation of the star is

\[ \Delta(c) = \max_{v,w \in V \setminus \{c\}} \delta(c,v,w) . \]
\end{defn}

\begin{defn}
\label{defn:mds}
Let 
\begin{itemize}
\item $P = \langle p_0, \ldots, p_{n-1} \rangle$ be a set of $n$ input
points from $\mathbb{R}^D$ for some constant $D$, each with the
potential to be the center of a star with vertices $P$,
\item $c_{OPT} \in \mathbb{R}^D$ be the point minimizing $\Delta(c_{OPT})$,
\item $S = \langle s_0, \ldots, s_{n-1} \rangle$ be the sequence
 formed by sorting $P$ by distance from $c_{OPT}$,
\item  $\epsilon>0$ be a constant parameter,
\item $\Gamma=2/\epsilon-1$,
\item $k$ be a constant depending only on $\Gamma$,
\item $L = \{ s_i \in S \enspace | \enspace 0 \leq i < k \}$ be the $k$-low centers,
\item and $H = P \setminus L$ be the $k$-high centers.
\end{itemize}
\end{defn}

We require the following claim, which is proved in \cite{ew05}:

\begin{clm}
\label{clm:annuli}
Let $c$ be the center of a Euclidean star in $\mathbb{R}^D$ for $D \in
O(1)$ having vertices $V$.  If $\Delta(c) \leq \Gamma$ for some
constant $\Gamma$, then there exists a constant $\rho_\Gamma$ such
that for any integer $i$, the $D$-dimensional annulus centered on $c$
with inner radius $\rho_\Gamma^i$ and outer radius $\rho_\Gamma^{i+1}$
contains only $O(1)$ points from $V$.
\end{clm}

\begin{lem}
\label{lem:constant-k}
For any $\Gamma$ there exists a constant $k$ depending only on
$\Gamma$ such that any center $s_i \in S$ with $i \geq k$ has dilation
$\Delta(s_i) \geq \Gamma$.
\end{lem}

\begin{proof}
We first consider the case when $\Delta(c_{OPT}) > \Gamma$.  By
definition $\Delta(s_i) \geq \Delta(c_{OPT})$ for any $s_i \in S$, so
we have that every $\Delta(s_i) > \Gamma$ regardless of the value of
$k$.

We now turn to the case when $\Delta(c_{OPT}) \leq \Gamma$.  Define
$\rho_\Gamma$ as in Claim \ref{clm:annuli}, and let
\[ A_j = \{ x \in \mathbb{R}^d \enspace | \enspace \rho_\Gamma^j \leq |x c_{OPT} | \leq \rho_\Gamma^{j+1} \} \]
\noindent be the $j$th annulus centered on $c_{OPT}$.  Define $l = 1 +
\lceil \log_{\rho_\Gamma} (\Gamma-1) \rceil$, and suppose $s_i \in
A_j$.  Let $v,w$ be two input points that lie in the annulus
$A_{j-l}$.  By the definition of $l$ we have

\begin{eqnarray*}
\rho_\Gamma^{l-1} - 1& \geq & \Gamma \\
\frac{2(\rho_\Gamma^{j-1} - \rho_\Gamma^{j-l})}{2 \rho_\Gamma^{j-l}} & \geq & \Gamma .
\end{eqnarray*}

\begin{figure}[t]
\centering
\includegraphics[height=3in]{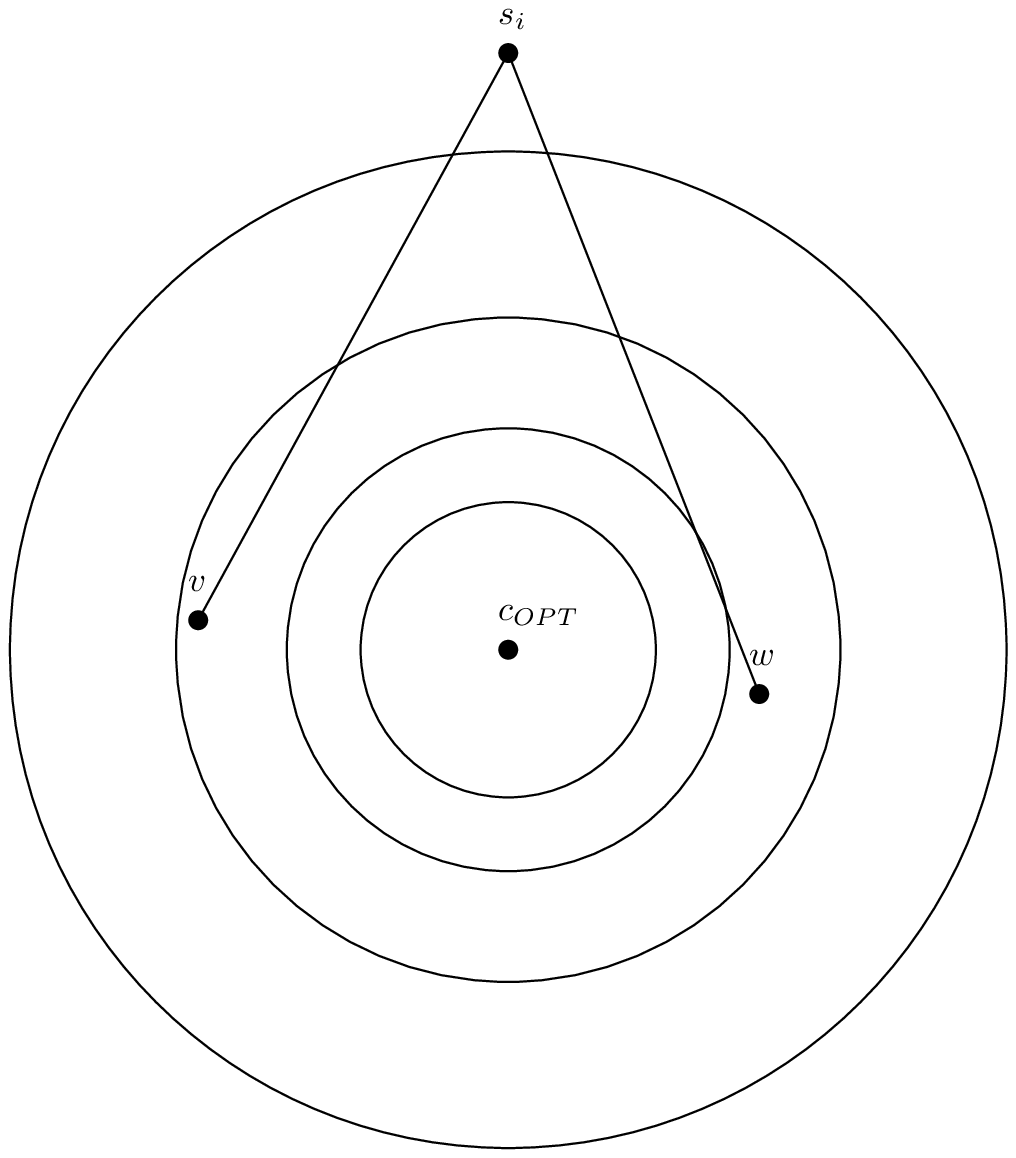}
\caption{Planar annuli containing the points $s_i$, $v$, and $w$.}
\label{figure:annuli}
\end{figure}

Since $v \in A_{j-l}$, we have $|s_i v| \geq (\rho_\Gamma^{j-1} -
\rho_\Gamma^{j-l})$, and similarly $|s_i w| \geq (\rho_\Gamma^{j-1} -
\rho_\Gamma^{j-l})$.  Further, $|vw| \leq 2 \rho_\Gamma^{j-l}$.  So by
substitution

\begin{eqnarray*}
\frac{(|s_i v|)+(|s_i w|)}{(|vw|)} & \geq & \Gamma \\
\Delta(s_i) & \geq & \Gamma .
\end{eqnarray*}

Thus the lemma holds if the annuli $A_j, A_{j-1}, \ldots, A_{j-l}$
contain at least $k$ points.  We can ensure this by selecting $k$ such
that the points $s_{i-k}, \ldots, s_i$ necessarily span at least $l+1$
annuli.  By Claim \ref{clm:annuli}, there exists some $m \in O(1)$
such that any annulus contains no more than $m$ input points.  So we
set $k \geq (l+1)m$.  The observation that $l$ depends only on
$\Gamma$ completes the proof.

\end{proof}

\begin{cor}
\label{cor:wrapup}
For any $c \in L$, $\Delta(c) \leq \Gamma$; for any $c \in H$,
$\Delta(c) \geq \Gamma$; and $|L| \in O(1)$.
\end{cor}

\begin{proof}
Each claim follows from Lemma \ref{lem:constant-k} and Definition
\ref{defn:mds}.
\end{proof}

\begin{lem}
\label{lem:low-time}
The set $L$, as well as the quantity $\Delta(c)$ for every $c \in L$,
may be computed in $O(n \log n)$ expected time.
\end{lem}

\begin{proof}
The unconstrained center $c_{OPT}$ may be found in $O(n \log n)$
expected time \cite{ew05}. $L$ may be constructed by sorting $P$ by
distance from $c_{OPT}$, and retaining the first $k$ elements, where
$k$ is the constant defined in Lemma \ref{lem:constant-k}.

The set $C$ may also be constructed in $O(n \log n)$ time \cite{ew05}.
Then any $\Delta(c)$ may be computed by evaluating

\[ \Delta(c) = \max_{(v,w) \in C} \delta(c,v,w), \]

\noindent
which takes $O(n)$ time since $|C| \in O(n)$ \cite{ew05}.  Evaluating
all $|L|=k$ dilation values this way takes $O(kn)$ time.  So the
expected amount of time needed to compute every $\Delta(c)$ is $O(n
\log n + kn)$, which is $O(n \log n)$ by Corollary \ref{cor:wrapup}.
\end{proof}

\begin{lem}
\label{lem:similar-legs}
If $c \in H$, $p_i, p_j \in P$ be the pair of points such that
$\delta(c,p_i,p_j)=\Delta(c)$, and $|p_i c| \geq |p_j c|$, then $|p_j
c| \geq (1-\epsilon) |p_i c|$.
\end{lem}

\begin{proof}
By the assumption $c \in H$ and Corollary \ref{cor:wrapup},
\[ \delta(c,p_i,p_j) = \frac{|p_i c| + |p_j c|}{|p_i p_j|} \geq \Gamma ; \]
by the triangle inequality
\begin{eqnarray*}
|p_i p_j| + |p_j c| & \geq & |p_i c| \\
|p_i p_j| & \geq & |p_i c| - |p_j c| ,
\end{eqnarray*}
\noindent so
\begin{eqnarray*}
\frac{|p_i c| + |p_j c|}{(|p_i c| - |p_j c|)} & \geq & \Gamma \\
|p_i c| + |p_j c| & \geq & \Gamma (|p_i c| - |p_j c|) \\
|p_j c| + \Gamma |p_j c| & \geq & \Gamma |p_i c| - |p_i c| \\
|p_j c| (\Gamma + 1) & \geq & |p_i c| (\Gamma - 1) \\
|p_j c| \geq \frac{\Gamma-1}{\Gamma+1} |p_i c| .
\end{eqnarray*}

By the definition of $\Gamma$,

\begin{eqnarray*}
|p_j c| & \geq & \frac{(2/\epsilon-1)-1}{(2/\epsilon-1)+1} |p_i c|\\
& \geq & \frac{2/\epsilon-2}{2/\epsilon} |p_j c|\\
& \geq & (1 - \epsilon) |p_j c| .\\
\end{eqnarray*}
\\
\end{proof}

\begin{lem}
\label{lem:approx-weight}
Define $q_{i,j} = \frac{1}{2}(p_i + p_j)$ and $w_{i,j} = \frac{2}{|p_i
p_j|}$; then for any $c \in H$ and corresponding $p_i, p_j \in P$ such
that $\delta(c,p_i,p_j)=\Delta(c)$,

\[ w_{i,j} \cdot |q_{i,j} c| \geq (1-\epsilon) \cdot \delta(c, p_i, p_j) .\]
\end{lem}

\begin{figure}[t]
\centering
\includegraphics[width=4in]{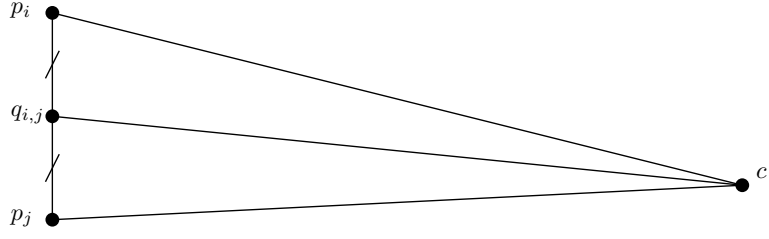}
\caption{Example points $p_i$, $p_j$, $q_{i,j}$, and $c$.}
\label{figure:triangle}
\end{figure}

\begin{proof}
By Lemma \ref{lem:similar-legs}
\[ |p_j c| \geq (1-\epsilon)|p_i c| \]
\noindent and by construction $|q_{i,j} c| \geq |p_j c|$, so
\begin{eqnarray*}
|q_{i,j} c| & \geq & (1-\epsilon) |p_i c| \\
2 \cdot |q_{i,j} c| & \geq & (1-\epsilon) (|p_i c| + |p_i c|) .\\
\end{eqnarray*}
By assumption $|p_i c| \geq |p_j c|$, so
\begin{eqnarray*}
2 \cdot |q_{i,j} c| & \geq & (1-\epsilon) (|p_i c| + |p_j c|) \\
\frac{2}{|p_i p_j|} |q_{i,j}c| & \geq & (1-\epsilon) \frac{|p_i c| + |p_j c|}{|p_i p_j|} ,
\end{eqnarray*}
and by substitution
\[ w_{i,j} \cdot |q_{i,j} c| \geq (1-\epsilon) \cdot \delta(c, p_i, p_j) .\]
\\
\end{proof}

\begin{thm}
\label{thm:mds-punchline}
A set of $n$ values $\hat{\Delta}(p_i)$ subject to
\[ (1-\epsilon) \Delta(p_i) \leq \hat{\Delta}(p_i) \leq \Delta(p_i) \]
may be computed in $O(n \log n)$ expected time.
\end{thm}

\begin{proof}
By Lemma \ref{lem:low-time} is is possible to separate the $k$-low and
$k$-high centers, and compute the exact dilation of all the $k$-low
centers, in $O(n \log n)$ expected time.  Section \ref{data-structure}
describes a data structure that can answer approximate weighted
farthest neighbor queries within a factor of $(1-\epsilon)$ in $O(\log
n)$ time after $O(n \log n)$ preprocessing.  As shown in Lemma
\ref{lem:approx-weight}, the dilation function $\delta$ may be
approximated for the $k$-high centers up to a factor of $(1-\epsilon)$
using weights $w_{i,j} = \frac{2}{|p_i p_j|}$.  So for each $k$-high
$p_i$, $\Delta(p_i)$ may be approximated by the result of a weighted
farthest neighbor query.  Each of these $|H|=n-k$ queries takes
$O(\log n)$ time, for a total expected running time of $O(n \log n)$.
\end{proof}

\begin{cor}
\label{cor:mds-hub}
Let $\Delta_{OPT} = \min_{p_i \in P} \Delta(p_i)$.  Then a point
$\hat{c} \in P$ satisfying
\[ (1-\epsilon) \Delta_{OPT} \leq \Delta(\hat{c}) \leq \Delta_{OPT} \]
may be identified in $O(n \log n)$ expected time.
\end{cor}

\begin{proof}
Theorem \ref{thm:mds-punchline} shows that the values
$\hat{\Delta}(p_i)$ may be computed in $O(n \log n)$ expected time.  A
suitable $\hat{c}$ may be found by generating these $n$ values, then
searching for the smallest $\hat{\Delta}(p_i)$ and returning the
corresponding $p_i$.
\end{proof}

% ----------------------------------------------------------------
\bibliographystyle{abbrv} \bibliography{awfn}

\begin{thebibliography}{1}

\bibitem{AgaHarVar04}
P.~K. Agarwal, S.~Har-Peled, and K.~R. Varadarajan.
\newblock Geometric approximation via coresets.
\newblock In J.~E. Goodman, J.~Pach, and E.~Welzl, editors, {\em Combinatorial
  and Computational Geometry}, number~52 in MSRI Publications, pages 1--30.
  Cambridge University Press, 2005.

\bibitem{am05}
S.~Arya and D.~Mount.
\newblock {Computational geometry: Proximity and location}.
\newblock In D.~P. Mehta and S.~Sahni, editors, {\em {Handbook of Data
  Structures and Applications}}, pages 63--1--63--22. CRC Press, 2005.

\bibitem{Cha04}
T.~M. Chan.
\newblock Faster core-set constructions and data stream algorithms in fixed
  dimensions.
\newblock In {\em Proc. 20th Symp. Computational Geometry}, pages 152--159, New
  York, NY, USA, 2004. ACM Press.

\bibitem{dg05}
C.~Duncan and M.~T. Goodrich.
\newblock {Approximate geometric query structures}.
\newblock In D.~P. Mehta and S.~Sahni, editors, {\em {Handbook of Data
  Structures and Applications}}, pages 26--1--26--17. CRC Press, 2005.

\bibitem{ew05}
D.~Eppstein and K.~A. Wortman.
\newblock Minimum dilation stars.
\newblock In {\em Proc. 21st Symp. Computational Geometry}, pages 321--326. ACM
  Press, Jun 2005.

\end{thebibliography}
\end{document}